\documentclass[a4paper,12pt]{article}

\usepackage{amssymb,amscd,amsmath,amsxtra,mathdots}
\long\def\del#1\enddel{}
\usepackage[english]{babel}
\usepackage{graphicx}
\usepackage{enumitem} 
\usepackage[font=small,labelfont=bf]{caption}

\usepackage{parskip}
\usepackage{color,soul,rotating}

\newcommand{\stirlingone}[2]{\genfrac{[}{]}{0pt}{}{#1}{#2}}

\def\E{{\mathcal E}}
\def\S{{\mathcal S}}

\def\J{{\mathcal J}}
\def\O{{\mathcal O}}

\def\ds{\displaystyle}
\def\r{\rho}
\def\a{\alpha}

\def\G{\Gamma}

\def\x{\xi}

\def\n{\nu}

\def\vf{\varphi}

\def\l{\lambda}

\def\o{\omega}
\def\sinh{\mathrm{sinh}}
\def\cosh{\mathrm{cosh}}

\def\p{\partial}
\def\rb{\right}
\def\lb{\left}

\def\cp{\mathbb {CP}^3}

\newcommand{\eq}[1]{\begin{equation} #1 \end{equation}}
\newcommand{\al}[1]{\begin{align} #1 \end{align}}

\newcommand{\bE}{\ensuremath{\mathbb{E}}}

\newcommand{\bK}{\ensuremath{\mathbb{K}}}

\numberwithin{equation}{section}

\begin{document}
\begin{titlepage}
\markright{\bf TUW--14--xx}
\title{Large J expansion in ABJM theory revisited}
\author{H.~Dimov${}^{\star}$\thanks{e-mail: h\_dimov@phys.uni-sofia.bg} ,\,
 S.~Mladenov${}^{\star}$ \, and R.~C.~Rashkov${}^{\dagger,\star}$\thanks{e-mail: rash@hep.itp.tuwien.ac.at}
\ \\ \ \\
${}^{\star}$ Department of Physics, Sofia University,\\
5 J. Bourchier Blvd, 1164 Sofia, Bulgaria
\ \\ \ \\
${}^{\dagger}$ Institute for Theoretical Physics,\\
Vienna University of Technology,\\
Wiedner Hauptstr. 8-10, 1040 Vienna, Austria
}
\date{}
\end{titlepage}

\date{}

\maketitle

\abstract{Recently there has been progress in the computation of the anomalous dimensions of gauge theory operators at strong coupling by making use of AdS/CFT correspondence. 
On string theory side they are given by dispersion relations in the semiclassical regime. We revisit the problem of large charges expansion of the dispersion relations for simple semiclassical strings in $AdS_4\times\mathbb{CP}^3$ background. We present the calculation of the corresponding
anomalous dimensions of the gauge theory operators to an arbitrary order using three different methods. Although the results of the three methods look different, power series expansions show their consistency.
}


\setcounter{tocdepth}{2}

\section{Introduction} 

Currently the world-volume dynamics of the branes is the most important player in 
the duality between gauge theories and strings (M-theory).

Last years there has been large number of works focused on the understanding of the world-volume dynamics of
multiple M2-branes and the near horizon limit of the effective geometry. This interest was inspired by the investigations of Bagger, Lambert and Gustavsson \cite{Bagger:2006sk} on theories having hidden structures of Lie 3-algebra and which have intimate relation to membrane dynamics. On other hand the progress in AdS/CFT correspondence motivates to look for a new class of conformal invariant, maximally supersymmetric
field theories in 2+1 dimensions describing the world-volume dynamics of coincident
membranes in M-theory. Indeed, a dual pair of theories corresponding to the above picture was found \cite{Schwarz:2004yj,Aharony:2008ug} and it triggered large number of investigations in various directions.
One side of the so called  Aharony-Bergman-Jafferis-Maldacena (ABJM) theory consist of
$N$ membranes on $S^7/\mathbb Z_k$ and membranes in M-theory
on $AdS_4\times S^7/\mathbb Z_k$, or after reduction on M-theory cycle to type IIA string theory, strings in $AdS_4\times \mathbb{CP}^3$ background. On the dual gauge theory side the theory is ${\cal N}=6$ superconformal Chern-Simons theory coupled with bi-fundamental matter (actually two Chern-Simons theories of level $k$ and 
$-k$ correspondingly, and each with gauge group $SU(N)$).  The superstrings on $AdS_4\times\cp$ as a coset was 
first studied in \cite{Arutyunov:2008if}\footnote{See also \cite{Stefanski:2008}} opening the door for investigation of the integrable structures in the theory. Shortly after that it was noticed that
the string supercoset model does not describe the entire dynamics of type IIA superstring in 
$AdS_4 \times \mathbb{CP}^3$, but only its subsector. 
The complete string dual of the ${\cal N}=6$ superconformal Chern-Simons theory,
 i.e. the complete type-IIA Green-Schwarz string action in  $AdS_4\times\cp$ superspace has been constructed in \cite{Gomis:2008jt}. 

Being highly non-linear, the theory on both sides is hard to solve exactly. Thus, the semiclassical
analysis appears to be the most appropriate available tool to answer many questions. 
The duality between the two theories suggests that the partition functions of string theory on
$AdS_4\times \mathbb{CP}^3$ and  ${\cal N}=6$ superconformal Chern-Simons theory are equal.
If one works on string side, one can find semiclassical the string solutions and workout string spectrum.
According to the AdS/CFT correspondence, the dispersion relations on string theory side are equal to the dimensions of the gauge theory operators. 
Therefore, one of the main ingredients necessary to check the holographic correspondence are the dimension of the gauge theory operators.

Although the issue of dispersion relations was addressed in some papers, see for instance\footnote{
There are huge amount of papers on ABJM theory. Here we will quote only those which are directly related to our study and which we really used.}
\cite{Minahan:2008hf}-\cite{Gomis:2008jt},\cite{Chen:2008qq,Rashkov:2008rm,Miramontes:2008wt},\cite{Arutyunov:2008if}-\cite{Ryang:2008rc}, in this note we revisit the problem studying the large momentum expansion of folded string dispersion relations. The difficulty is that the conserved charges for corresponding semiclassical string solutions are typically represented in terms of elliptic integrals. The latter is difficult to invert separating carefully leading, sub-leading etc. orders of the contributions. In this note we consider three methods for calculating the anomalous dimensions of the gauge theory operators by AdS/CFT correspondence. The result agree in sense that they give the same results, but in quite different form. 

This paper is organized as follows. In the Introduction we give very brief review of the basics of ABJM theory and review a simple folded string solution in $\mathbb{CP}^3$ part of the geometry. Next we apply the iterative method for inverting the elliptic integrals to obtain the dispersion relations corresponding to the folded string solution.
The second method used for calculation of the dispersion relations is the Picard-Fuchs equation. Finally we present the calculation of the anomalous dimensions to the first a few orders using Lambert function as advocated in 
\cite{Floratos:2013cia} for the case of $AdS_5\times S^5$.

\subsection{On ABJM theory in brief}

The $AdS_4/CFT_3$ correspondence, or ABJM theory, is 
one of the rare candidates for exact string/gauge theory correspondence.
It is obtained starting from 11d M-theory analysing M2-brane
dynamics. One starts with M2-brane solutions obtained considering  11 dimensional 
supergravity action \cite{Aharony:2008ug} 
\eq{
S=\frac{1}{2\kappa_{11}^2}\int
dx^{11}\sqrt{-g}\left(R-\ds\frac{1}{2\cdot
4!}F_{\mu\nu\rho\sigma}F^{\mu\nu\rho\sigma}\right)-\frac{1}{12\kappa_{11}^2}
\int C^{(3)}\wedge F^{(4)}\wedge F^{(4)}, \label{abjm-1} 
}
 where $\kappa_{11}^2=2^7\pi^8 l_p^9$. The M2-brane solutions can be obtained from 
 11d SUGRA equations of motions,
\eq{
R^\mu_\nu=\frac{1}{2}\left(\frac{1}{3!}F^{\mu\a\beta\gamma}F_{\nu\a\beta\gamma}
-\frac{1}{3\cdot 4!}\delta^\mu_\nu
F_{\a\beta\rho\sigma}F^{\a\beta\rho\sigma}\right), \label{abjm-2}
} 
and 
\eq{
\p_{\sigma}(\sqrt{-g}F^{\sigma\mu\nu\xi})=\frac{1}{2\cdot
(4!)^2}\epsilon^{\mu\nu\xi\a_1\dots\a_8}F_{\a_1\dots\a_4}F_{\a_5\dots\a_8}.
\label{abjm-3} 
}
For the purpose of AdS/CFT correspondence we are interested in near horizon limit of $AdS_4\times S^7$ spacetime
\eq{
ds^2=\frac{R^2}{4}ds^2_{AdS_4}+R^2 ds^2_{S^7}, \label{abjm-4} 
} 
where the $AdS_4$ cycle supports  $N'$ units of four-form flux 
\eq{
 F^{(4)}=\frac{3R^3}{8}\epsilon_{AdS_4}, \quad R=l_p(2^5 N'\pi^2)^{\frac{1}{6}}.
\label{abjm-5} 
}

Let us consider the quotient $S^7/\mathbb Z_k$ where $\mathbb{Z}_k$ is acting as 
$z_i \rightarrow e^{ i \frac{2 \pi}{ k} } z_i$. A convenient way to proceed is 
 write the metric on $S^7$ as
\eq{ 
ds^2_{S^7} =  ( d \varphi' + \omega)^2 + ds^2_{CP^3},
\label{abjm-6} 
} 
where 
\al{ 
& ds^2_{CP^3} = \frac{ \sum_i d z_i d \bar z_i}{r^2 } - \frac{ | \sum_i z_i d \bar z_i |^2}{ r^4} ~,\quad
 r^2 \equiv \sum_{i=1}^4 |z_i|^2,\notag \\
& d \varphi' + \omega  \equiv  \frac{ i}{2  r^2 } (  z_i d \bar z_i - \bar z_i d z _i ),
\quad d\omega =  J = { i}   d \left(\frac{ z_i}{r}\right)  d \left(\frac{ \bar z_i}{ r }\right),
\label{abjm-7} 
} 
and then perform the $\mathbb Z_k$ quotient identifying $\vf'=\vf/k$ with $\vf\sim \vf+2\pi$. 
Noticing that $J$ is proportional to the K\"{a}hler form on $\cp$ one can write resulting metric as 
\eq{ 
ds^2_{S^7/{\mathbb Z}_k} = \frac{ 1}{k^2 } ( d \vf+ k \o)^2 + ds^2_{CP^3}. 
} 
The consistent quantization of the flux forces the condition $N'=kN$, where $N$ is the number of quanta of the flux on the quotient. In this way the spectrum of the resulting theory will be that of the initial $AdS_4\times S^7$ projected onto the  $\mathbb Z_k$ invariant sector.  
In this setup there is a natural definition of {}'t Hooft
coupling $\lambda \equiv N/k$ and the decoupling limit is 
$N,k\rightarrow \infty$ while $N/k$ is kept fixed \cite{Aharony:2008ug}.

On gauge theory side it was conjectured that N multiple M2-
branes on $\mathbb{C}^4/\mathbb{Z}_k$ is described by $\mathcal{N} = 6$
 supersymmetric Chern-Simons-matter theory with
gauge group $U(N) \times U(N)$ and levels $k$ and $-k$ correspondingly.

The Chern-Simons part of the theory is constructed using a pair of chiral fields $A_i$ (i = 1, 2)
in the bifundamental representation $(\mathbf{N},\bar{\mathbf{N}})$, and
a pair $B_i$ in the anti-bifundamental representation $(\bar{\mathbf{N}},\mathbf{N})$. 
The theory is supplied with  a $\mathcal{N} = 2$ superpotential
\eq{
W =\frac{4\pi}{k}\mathrm{Tr}(A_1B_1A_2B_2 - A_1B_2A_2B_1),
}
where the scalar components of $(A_1,A_2,B^\dagger_1,B^\dagger_2)$ transform in the 
$\mathbf{4}$ of $SU(4)_R$ and the conjugate in the $\bar{\mathbf{4}}$.
The  $\mathcal{N} = 6$ supersymmetry is combined with an exact conformal symmetry 
organized in  $OSp(6|4)$  superconformal symmetry group. 

\paragraph{ABJM and strings on $AdS_4\times \cp$}

One can follow now \cite{Aharony:2008ug} to make reduction to type IIA  with
the following final result 
\al{
 ds^2_{string} = & \frac{ R^3}{ k} \left( \frac{ 1 }{ 4 } ds^2_{AdS_4} + ds^2_{{\mathbb CP}^3 } \right), \\
 e^{2 \phi} = & \frac{ R^3 }{ k^3 } \sim \frac{ N^{1/2} }{ k^{5/2} }= \frac{ 1 }{ N^2 } \left( 
 \frac{N }{ k } \right)^{5/2}, \\
 F_{4} = & \frac{ 3 }{ 8 }  {  R^3}  \epsilon_4 , \quad F_2 =  k d \omega = k J,
 }
 We end up then with  $AdS_4 \times\cp$ compactification
 of type IIA string theory with $N$ units of $F_4$ flux on $AdS_4$ and $k$ units of
$F_2$ flux on the ${\mathbb CP}^1 \subset\cp$ 2-cycle.

The radius of curvature in string units is $R^2_{str} = 
\frac{ R^3}{ k}  =  2^{5/2} \pi \sqrt{ \lambda}$. It is important to note
that the type IIA approximation is valid in the regime where $k
\ll N \ll k^5$.


To proceed, we will need the explicit form of the metric on $AdS_4\times\mathbb{CP}^3$ in spherical coordinates. The convenient form of the
metric on $AdS_4\times\mathbb{CP}^3$ can be written as
\cite{PopeWarner}
\begin{multline}
 ds^2=R^2\left\lbrace \dfrac{1}{4}\left[ -\cosh^2\rho\,dt^2+d\rho^2+\sinh^2\rho\,d\Omega_2^2\right]+d\mu^2
  \right.\\
 \left.+\sin^2\mu\left[d\alpha^2+\dfrac{1}{4}\sin^2\alpha(\sigma_1^2
+\sigma_2^2+\cos^2\alpha\sigma_3^2)+\dfrac{1}{4}\cos^2\mu(d\chi+\sin^2\mu\sigma_3)^2\right]\right\rbrace.
\label{metric}
\end{multline}
Here $R$ is the radius of the $AdS_4$, and $\sigma_{1,2,3}$ are
left-invariant 1-forms on an $S^3$, parameterized by
$(\theta,\phi,\psi)$,
\begin{align}
&\sigma_1=\cos\psi\,d\theta+\sin\psi\sin\theta\,d\phi,\notag\\
&\sigma_2=\sin\psi\,d\theta-\cos\psi\sin\theta\,d\phi,\label{S3}\\
&\sigma_3=d\psi+\cos\theta\,d\phi.\notag
\end{align}
The range of the coordinates is
$$0\leq\mu,\,\alpha\leq\dfrac{\pi}{2},\,\,0\leq\theta\leq\pi,\,\,
0\leq\phi\leq2\pi,\,\,0\leq\chi,\,\psi\leq4\pi.$$

\subsection{Review of the simplest semiclassical string solutions in $AdS_4\times\mathbb{CP}^3$ background}

As we noticed above the Chern-Simons terms are with opposite sign levels, namely $k$ and $-k$, while the superpotential has an $SU(2) \times SU(2)$ global symmetry  acting on the pairs $A_i$ and  $B_i$
correspondingly. In the near horizon limit with $k$ and $N$ satisfying 
$k \ll N \ll k^5$ the field theory
is dual to IIA superstring theory on $AdS_4 \times \mathbb{CP}^3$ with constant dilaton, RR two-form and
four-form fluxes. 
To make the symmetries of the background explicit let us write the metric and the field content in the form
\begin{align}
& ds^2_{IIA}=\frac{R^3}{k}\lb(\frac{1}{4}
ds^2_{AdS_4}+ds^2_{\cp}\rb) \notag \\
& ds^2_{AdS_4}=R^2(-\cosh^2\rho dt^2 + d\rho^2 + \sinh^2\rho(d\theta^2 + \sin^2\theta d\phi^2))\notag \\
& ds^2_{\cp}=d\xi^2+\cos^2\xi\sin^2\xi\lb(d\psi+\frac{1}{2}\cos\theta_1d\vf_1-
\frac{1}{2}\cos\theta_2d\vf_2\rb)^2 \notag \\
& \qquad +\frac{1}{4}\cos^2\xi(d\theta_1^2+\sin^2\theta_1 d\vf_1^2)
 +\frac{1}{4}\sin^2\xi(d\theta_2^2+\sin^2\theta_2 d\vf_2^2),\notag \\
& C_1=\frac{k}{2}\lb((\cos^2\xi-\sin^2\xi)d\psi+\cos^2\xi\cos\theta_1d\vf_1+\sin^2\xi\cos\theta_2
d\vf_2\rb), \notag \\
& F_2=k\lb(-\cos\xi\sin\xi\,d\xi\wedge(2d\psi+\cos\theta_1d\vf_1-\cos\theta_2d\vf_2)
\right. \notag \\
&\left.\qquad -\frac{1}{2}\cos^2\xi\sin\theta_1d\theta_1\wedge d\vf_1-\frac{1}{2}\sin^2\xi
\sin\theta_2 d\theta_2\wedge d\vf_2 \rb) \notag \\
& F_4=-\frac{3R^3}{8}\o_{AdS_4},\quad  e^{2\Phi}=\frac{R^3}{k^3} \label{metric-fields}.
\end{align}
Written in this form, it is easy to see that the background has at least five killing vectors corresponding
to the translations along $t,\psi,\phi,\varphi_1,\varphi_2$. The charges associated with the killing vectors
are the energy and the momenta $S,J_1,J_2,J_3$.

To this end it is plausible to make an ansatz for this directions \cite{Chen:2008qq}:
\begin{equation}
t=\kappa \tau, \quad \phi=v\tau, \quad \psi=\omega_1\tau, \quad \varphi_1=\omega_2\tau,
\quad \varphi_2=\omega_3\tau.
\label{base-ansatz}
\end{equation}
In  this setup the authors of  \cite{Chen:2008qq} found a simple classical string solutions and the corresponding charges. The latter are defined as:
\begin{align}
& E=\frac{1}{4}\cosh^2\rho\sqrt{\tilde{\lambda}}\kappa\notag \\
& S=\sqrt{\tilde{\lambda}}\int\frac{d\sigma}{2\pi}v\sinh^2\rho\sin^2\theta, \notag \\
& J_1=\sqrt{\tilde{\lambda}}\int\frac{d\sigma}{2\pi}\sin^2\xi\cos^2\xi
\left(\omega_1+\cos\theta_1\frac{\omega_2}{2}-\cos\theta_2\frac{\omega_3}{2} \right) \\
& J_2=\sqrt{\tilde{\lambda}}\int\frac{d\sigma}{2\pi}\left[
\cos^2\xi\sin^2\theta_1\frac{\omega_2}{4}+\cos^2\xi\sin^2\xi
\left(\cos^2\theta_1\frac{\omega_2}{4}+\cos\theta_1(\frac{\omega_1}{2}-\cos\theta_2\frac{\omega_3}{4})\right)
\right] \notag \\
& J_3=\sqrt{\tilde{\lambda}}\int\frac{d\sigma}{2\pi}\left[
\cos^2\xi\sin^2\theta_2\frac{\omega_3}{4}+\cos^2\xi\sin^2\xi
\left(\sin^2\theta_2\frac{\omega_3}{4}-\cos\theta_2(\frac{\omega_1}{2}+\cos\theta_1\frac{\omega_2}{4})\right)
\right]\notag
\end{align}
The relation between $\tilde{\lambda}$ and the {}'t Hooft coupling is $\tilde{\lambda}=
\sqrt{32}\pi\lambda$.

If we restrict ourselves to the case of strings moving in $\mathbb{R}_t \times \mathbb{CP}^3$,
 the angular momenta in one $S^2$ are opposite to those in the other $S^2$ while executing
motion on $S^1$ in the $U(1)$ Hopf fibration over $S^2 \times S^2$.
On the gauge theory side the BPS state corresponding to the string vacuum is
$\mathrm{tr}[(A_1B_1)^L]$. If we consider for instance the case of strings with two angular momenta, the corresponding composite states should be $\mathrm{tr}[(A_1B_1)^{J_1}(A_2B_2)^{J_2}]+perm$.

\paragraph{A simple semiclassical solution}

Below we briefly review the simplest folded string solution of \cite{Chen:2008qq}. We will work out in details 
 this case while the results for the more complicated solution are given in an Appendix.  To find  a simple folded sting solution, and in addition to \eqref{base-ansatz}, let us make the ansatz $\theta_1=\theta_2=0$. 
 The equation of motion for $\xi$ then take the following form
\begin{equation}
\xi''=\frac{1}{4}\sin4\xi\,\tilde{\omega}^2,
\end{equation}
where $\tilde{\omega}=\omega_1+(\omega_2-\omega_3)/2$. The Virasoro constraint is
\begin{equation}
\frac{\kappa^2}{4}=\xi'^2+\frac{\sin^2 2\xi}{4}\tilde{\omega}^2.
\end{equation}
Because we looking for a folded string here, $\xi$ will reach its maximal value at some $\xi_0$. At this point $\xi=\xi_0$ we have $\xi'=0$ and therefore $\kappa^2=\sin^2 2\xi_0\,\tilde{\omega}^2$. Now the Virasoro constraint have very simple form
\begin{equation}
\xi'^2=\frac{\tilde{\omega}^2}{4}\left(\sin^2 2\xi_0-\sin^2 2\xi\right),
\end{equation}
which is easy to solve in terms of elliptic Jacobi functions.
After integration of the above equation from the origin to the turning point we obtain the periodicity condition
\begin{equation}
2\pi=4\int_{0}^{\xi_0}\frac{2d\xi}{\tilde{\omega}\sqrt{\sin^2 2\xi_0-sin^2 2\xi}}.
\label{folding-cond}
\end{equation}
The angular momenta with this ansatz simply reduce to
\begin{equation}
J_1=\sqrt{\tilde{\lambda}}\int\frac{d\sigma}{2\pi}\cos^2 \xi \sin^2 \xi\,\tilde{\omega}
\end{equation}
\begin{equation}
J_2=\frac{J_1}{2}
\end{equation}
\begin{equation}
J_3=-J_2
\end{equation}
These relations show that the only independent charges are the energy and one of the angular momenta, say $J_1$.
Solving for $\xi$ with the condition \eqref{folding-cond} imposed, we obtain that energy and angular momenta of the folded string can be expressed in terms of linear combinations of complete elliptic integrals of first kind $\bK(k)$ and second kind $\bE(k)$ with modular parameter $k=\sin 2\xi_0$
\begin{equation}
E=\frac{\sqrt{\tilde{\l}}}{4}\kappa=\frac{\sqrt{\tilde{\l}}}{4}\tilde{\o}\sin 2\x_0=\frac{\sqrt{\tilde{\l}}}{2\pi}k\bK(k)
\label{energy}
\end{equation}
\begin{equation}
J_1=\frac{\sqrt{\tilde{\l}}}{2\pi}\left[\bK(k)-\bE(k)\right].
\label{momentum}
\end{equation}

\section{The dispersion relations}

Although there are many interesting developments and applications of this duality, some 
old questions about the main players in the story are still interesting. For instance, the dispersion relations which are supposed to give the anomalous dimensions are given in implicit form. Most frequent cases are those when the charges are expressed through elliptic integrals which cannot be inverted in closed form. We revisit this problem combining a few approaches to obtain the dispersion relation as series (computable to arbitrary order).

We will approach the problem expanding the expressions for the charges and inverting the series. 
Expansions for the elliptic integrals in series are given in Appendix \ref{elliptic}. 

Another approach we will use is finding recurrence for the coefficients using a specific Picard-Fuchs equation.
The latter is based on the following simple facts. It is a simple exercise to find that the complete elliptic integrals satisfy the equation
\begin{equation}
\frac{d\bK}{dk}=\frac{1}{kk^{\prime 2}}(\bE-k^{\prime 2}\bK), \quad \frac{d\bE}{dk}=
\frac{1}{k}(\bE-\bK).
\end{equation}
After differentiating the above equations we find a special form of the Picard-Fuchs equation
\begin{equation}
kk^{\prime 2}\frac{d^2\bK}{dk^2}+(1-3k^2)\frac{d\bK}{dk}-k\bK=0.
\end{equation}

The general derivation of the dispersion relations here follows that in  \cite{Pawellek:2011xd}. They however are derived for spinning strings in $AdS_5$ part of the geometry. This means that they describe gauge theory operators with $S$ impurities. We will see latter that, surprisingly, the case of rotating strings in $\mathbb{CP}^3$ part of our theory can be manipulated in a very similar way.
For completeness, let us give a brief review of the case considered in \cite{Pawellek:2011xd}. The energy and the spin in the case of spinning string in $AdS_5$ are
\begin{equation}
\E=\frac{2}{\pi}\frac{k}{1-k^2}\bE, \qquad \S=\frac{2}{\pi}\left[\frac{1}{1-k^2}\bE-\bK\right].
\end{equation}
Applying the above actions, we find
\begin{equation}
\frac{d\E}{dk}=\frac{1}{1-k^2}\left[\S+\frac{1}{k}\E\right], \quad \frac{d\S}{dk}=
\frac{1}{1-k^2}\left[k\S+\E\right].
\end{equation}
Note that
\begin{equation}
\frac{d\E}{d\S}=\frac{1}{k}.
\label{eq-1}
\end{equation}
One can proceed with the second derivative
\begin{equation}
\frac{d^2\E}{d\S^2}=\frac{d}{d\S}\left(\frac{d\E}{d\S}\right)=
\left(\frac{d\S}{dk}\right)^{-1}\frac{d}{dk}\left(\frac{1}{k}\right)\\
=-\frac{1-k^2}{k^2}\frac{1}{k\S+\E},
\end{equation}
and using \eqref{eq-1} we find
\begin{equation}
\label{fin-equation}
\left(\S+\E(\S)\frac{d\E}{d\S}\right)\frac{d^2\E}{d\S^2}+\left(\frac{d\E}{d\S}\right)^3
-\frac{d\E}{d\S}=0.
\end{equation}
Now one can expand for large $\S$ and find terms in the expansion. We will apply this approach to our case.

There are two kinds of folded strings- short and long ones. They are distinguished by the modular parameter of the elliptic integrals, small or close to 1 correspondingly, and therefore the expansion in the two cases is quite different.

Before we start the analysis of the two kind of approximations for the string solutions, namely the short and long folded string, let us make a comment on that.
There exists a remarkable duality between the values of conserved charges of the two extreme types of the folded strings, that is short strings ($\sin 2\xi_0\rightarrow 0$) and long strings ($\sin 2\xi_0\rightarrow 1$). The duality is a direct consequence of the Legendre relation that connects complete elliptic integrals of the first and second kind, precisely:
\begin{equation}\label{eq:Legendre}
\mathbb{E}(k)\mathbb{K}(k')+\mathbb{K}(k)\mathbb{E}(k')+\mathbb{K}(k)\mathbb{K}(k')=\frac{\pi}{2},
\end{equation}
where complementary elliptic modulus $k'$ is defined as $k^2+k'^2=1$ and $k=\sin 2\xi_0$. 
By solving \eqref{energy} and \eqref{momentum} for $\mathbb{E}(k)$ and $\mathbb{K}(k)$ and substituting in \eqref{eq:Legendre}, we get the following duality relation:
\begin{equation}\label{eq:EJduality}
\frac{1}{kk'}EE'-\frac{1}{k}EJ_1'-\frac{1}{k'}E'J_1=\frac{\tilde{\lambda}}{8\pi}.
\end{equation}
When $k\sim 1$ we have $k'\sim 0$ and vice versa. Formula \eqref{eq:EJduality} defines map between energies and angular momenta of short and long strings. Furthermore, it can be rewritten in terms of anomalous dimensions $\gamma=E-J_1$ as:
\begin{equation}\label{eq:Egammaduality}
\frac{1}{k'}E'\gamma+\frac{1}{k}E\gamma'+\left(\frac{1}{kk'}-\frac{1}{k}-\frac{1}{k'}\right)EE'=\frac{\tilde{\lambda}}{8\pi}.
\end{equation}

\paragraph{Expansion of the conserved charges for short folded strings}
The expansion for the case of short strings is easy to obtain. Here we give the result for completeness.
The energy and angular momentum expansions for short folded strings in terms of $k=\sin 2\xi_0$ assume the following forms:
\begin{equation}\label{eq:Eshort}
E=\frac{\sqrt{\tilde{\lambda}}}{4}\sum_{n=0}^\infty\left(\frac{(2n-1)!!}{(2n)!!}\right)^2k^{2n+1}
\end{equation}
\begin{equation}\label{eq:Jshort}
J_1=\frac{\sqrt{\tilde{\lambda}}}{4}\sum_{n=0}^\infty\left(\frac{(2n-1)!!}{(2n)!!}\right)^2\frac{2n}{2n-1}k^{2n}.
\end{equation}
Since the energy and angular momentum represent power series of $x=k^2$, the series \eqref{eq:Jshort} can be easily inverted either by hand or using symbolic computational program. Then the inverse spin function $x=x(\mathcal{J}_1)$ may be inserted into \eqref{eq:Eshort} which lead us to the dispersion relation $\mathcal{E}=\mathcal{E}(\mathcal{J}_1)$. The results are:
\begin{equation}
x=\frac{4\mathcal{J}_1}{\pi}-\frac{6\mathcal{J}_1^2}{\pi^2}+\frac{3\mathcal{J}_1^3}{\pi^3}+\frac{5\mathcal{J}_1^4}{4\pi^4}-\frac{9\mathcal{J}_1^5}{16\pi^5}-\frac{21\mathcal{J}_1^6}{16\pi^6}-\frac{35\mathcal{J}_1^7}{64\pi^7}+\frac{459\mathcal{J}_1^8}{512\pi^8}+\frac{5835\mathcal{J}_1^9}{4096\pi^9}+\ldots
\end{equation}
\begin{equation}
\mathcal{E}=\pi^{1/2}\mathcal{J}_1^{1/2}+\frac{\mathcal{J}_1^{3/2}}{4\pi^{1/2}}+\frac{3\mathcal{J}_1^{5/2}}{32\pi^{3/2}}+\frac{\mathcal{J}_1^{7/2}}{128\pi^{5/2}}-\frac{61\mathcal{J}_1^{9/2}}{2048\pi^{7/2}}-\frac{201\mathcal{J}_1^{11/2}}{8192\pi^{9/2}}+\frac{199\mathcal{J}_1^{13/2}}{65536\pi^{11/2}}+\ldots
\end{equation}
Going back to the dimensional energy and angular momentum, the latter may also be written in the form:
\begin{equation}
E=\left(\frac{\sqrt{\tilde{\lambda}}J_1}{2}\right)^{1/2}\left[1+\frac{J_1}{2\tilde{\lambda}^{1/2}}+\frac{3J_1^2}{8\tilde{\lambda}}+\frac{J_1^3}{16\tilde{\lambda}^{3/2}}-\frac{61J_1^{4}}{128\tilde{\lambda}^2}-\frac{201J_1^{5}}{256\tilde{\lambda}^{5/2}}+\mathcal{O}\left(\frac{J_1^6}{\tilde{\lambda}^3}\right)\right].
\end{equation}

\subsection{Expansion of the conserved charges for long folded strings}\label{sec:ExpLongFoldedStr}

For convenience, we define the following quantities
\begin{equation} \label{eq:1}
\E=\frac{2\pi}{\sqrt{\tilde{\l}}}E=k\bK, \qquad \J_1=\frac{2\pi}{\sqrt{\tilde{\l}}}J_1=\bK(k)-\bE(k).
\end{equation}

\begin{figure}
\centering
\includegraphics[scale=0.6]{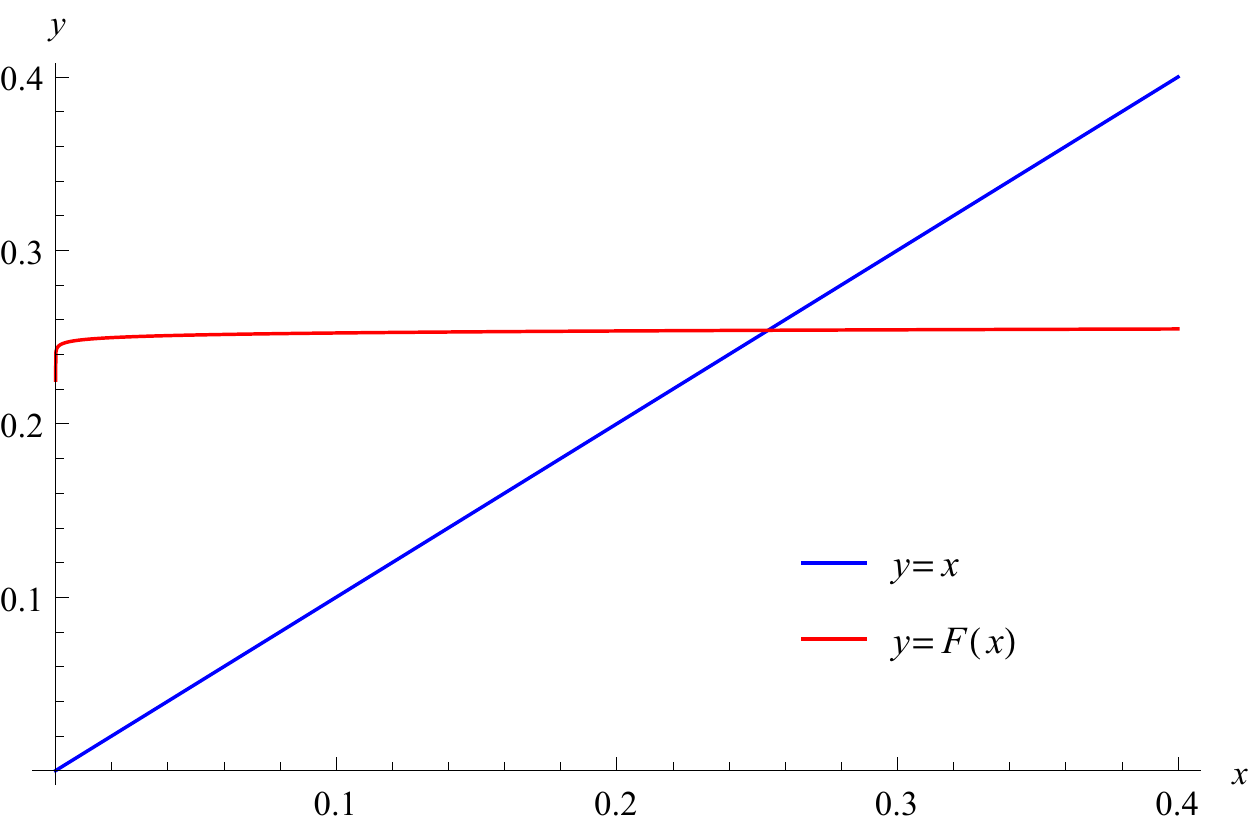}
\caption{The iteration function $F(x)$.}
\label{fig:IterationMap}
\end{figure}

Now let's try to expand the elliptic integrals and reverse the series following the ideas of \cite{Georgiou:2010zt}. 
Using formulas \eqref{eq:expansionK} and \eqref{eq:expansionE} from the Appendix \ref{elliptic} we can represent the energy and the spin \eqref{eq:1} in the form suitable for expansion:
\begin{equation}\label{eq:energy1}
\E=\sqrt{1-x}\sum_{n=0}^{\infty}x^n(a_n\ln x+b_n)
\end{equation}
\begin{align} \label{eq:spin1}
\J_1&=-1+\sum_{n=0}^{\infty}x^n(a_n\ln x+b_n)+\sum_{n=0}^{\infty}x^{n+1}(g_n\ln x+h_n)\nonumber\\
&=-1+b_0+a_0\ln x+\sum_{n=0}^{\infty}x^{n+1}(c_n\ln x+d_n)
\end{align}
where $x=1-k^2$ and
\begin{align}
&a_n=-\frac{1}{2^{2n+1}}\left[\frac{(2n-1)!!}{n!}\right]^2\nonumber\\
&b_n=a_n\left[2\sum_{k=1}^n\frac{1}{k(2k-1)}-4\ln 2\right]\nonumber\\
&c_n=a_{n+1}+g_n=\frac{1}{2^{2n+3}}\frac{(2n+1)!!(2n-1)!!}{[(n+1)!]^2}=-a_{n+1}\frac{1}{2n+1}\nonumber\\
&d_n=b_{n+1}+h_n=c_n\left[2\sum_{k=1}^{n+1}\frac{1}{k(2k-1)}+\frac{2}{2n+1}-4\ln 2\right].
\end{align}
In formula \eqref{eq:spin1} we separated the term $n=0$ in the first sum and then changed the index as follows $n\rightarrow n+1$. The explicit values of the first few coefficients are
\begin{align}
&a_0=-\frac{1}{2},\qquad &a_1=-\frac{1}{8},\qquad &a_2=-\frac{9}{128},\nonumber\\
&b_0=2\ln 2,\qquad &b_1=-\frac{1}{4}+\frac{1}{2}\ln 2,\qquad &b_2=-\frac{21}{128}+\frac{9}{32}\ln 2,\nonumber\\
&c_0=-\frac{1}{8},\qquad &c_1=\frac{3}{128},\qquad &c_2=\frac{5}{512},\nonumber\\
&d_0=\frac{1}{2}-\frac{1}{2}\ln 2,\qquad &d_1=\frac{9}{128}-\frac{3}{32}\ln 2,\qquad &d_2=\frac{43}{1536}-\frac{5}{128}\ln 2.
\end{align}
In order to reverse the series we represent them in a manner so that they look similar to one another. To accomplish that intention we have to expand the square root and collect the terms in front of $x$ up to $x^2$. Then we shall notice that it is more appropriate not to reverse the series individually one by one but their linear combination. The most convenient linear combination seems to be $\E_1=\J_1-\E+1$. After some simple algebra we reach to the following form $\E_1$:
\begin{equation}
\E_1=x(a_{10}\ln x+b_{10})+x^2(a_{20}\ln x+b_{20})+\O(x^3),
\end{equation}
where
\begin{align}
&a_{10}=c_0+\frac{a_0}{2}-a_1=0,\nonumber\\
&b_{10}=d_0+\frac{b_0}{2}-b_1=\frac{3}{4},\nonumber\\
&a_{20}=c_1+\frac{a_1}{2}+\frac{a_0}{8}-a_2=-\frac{1}{32},\nonumber\\
&b_{20}=d_1+\frac{b_1}{2}+\frac{b_0}{8}-b_2=\frac{7}{64}+\frac{1}{8}\ln 2,\nonumber\\
\end{align}
Reversion of the modified spin function $x=x(\E_1)$ passes through definition of the function $x^*(\E_1)$ as a solution of the ''reduced'' equation
\begin{equation}
\E_1=x^*b_{10}+x^{*2}(a_{20}\ln x^*+b_{20}).
\end{equation}
The function $x^*(\E_1)$ can be found by iteration of the following map (figure \ref{fig:IterationMap}):
\begin{equation}
F(x)=\frac{\E_1}{b_{10}}-\frac{x^2}{b_{10}}(a_{20}\ln x+b_{20}).
\end{equation}
The iteration $x_n=FF\dots F(x_0)$ starts from $x_0=\E_1$ and gives
\begin{align}
&x_0=\E_1\nonumber\\
&x_1=\frac{\E_1}{b_{10}}\cdot (1-\E_1A)\nonumber\\
&x_2=\frac{\E_1}{b_{10}}\cdot (1-\E_1A)\cdot\left[1+\sum_{i=1}^n\E_1^nA^n-\frac{\E_1}{b_{10}^2}\left(1-\E_1A\right)\left(A-a_{20}\ln b_{10}-\sum_{i=1}^n\frac{\E_1^nA^n}{n}\right)\right]\nonumber\\
&\vdots\nonumber\\
&x^*
\end{align}
where $A=a_{20}\ln\E_1+b_{20}$. The above infinite product representation of the iteration procedure is very convenient for computing logarithms of $x^*$. It is clear now that if we want to determine only the unknown constant term and the coefficient in front of $\ln\E_1$, we have to take just the zero iteration $x_0$.
\begin{align}
&\E=a_0\ln x+b_0+\O(x)=a_0\ln\E_1-a_0\ln b_{10}+b_0+\O(\E_1)\nonumber\\
&f=-\frac{1}{2},\qquad f_c=b_0-a_0\ln b_{10}=\ln 2+\frac{1}{2}\ln 3
\end{align}
Finally, one can write down the expansion of the energy $E$ with all the coefficients determined.
\begin{align}
&E(E_1)=\r_c+\r\ln(E_1)+\sum_{n=1}^{\infty}\sum_{m=0}^{n}\r_{nm}E_1^n\ln^mE_1,\nonumber\allowdisplaybreaks[4]\\
&\r=-\frac{\sqrt{\tilde{\l}}}{4\pi},\qquad\r_c=\frac{\sqrt{\tilde{\l}}}{4\pi}\ln\frac{6\sqrt{\tilde{\l}}}{\pi},\nonumber\allowdisplaybreaks[4]\\
&\r_{11}=\frac{1}{4},\qquad\r_{10}=-\frac{1}{4}\left(\frac{3}{2}+\ln\frac{6\sqrt{\tilde{\l}}}{\pi}\right),\allowdisplaybreaks[4]\\
&\r_{22}=\frac{\pi}{8\sqrt{\tilde{\l}}},\qquad \r_{21}=\frac{\pi}{8\sqrt{\tilde{\l}}}\left(1-2\ln\frac{6\sqrt{\tilde{\l}}}{\pi}\right),\nonumber\allowdisplaybreaks[4]\\
&\r_{20}=\frac{\pi}{8\sqrt{\tilde{\l}}}\left(\frac{5}{4}-6\ln2-3\ln3+\ln^2 12+\ln^2\frac{\sqrt{\tilde{\l}}}{2\pi}\right).\nonumber
\end{align}

\paragraph{Expansion using Picard-Fuchs equations.}

Let us see now how the idea of using Picard-Fuchs equations works in our case.
Since \eqref{eq:1} are additive combinations (see Appendix \ref{elliptic}) of the complete elliptic integrals they inherit the property that their derivatives with respect to modular parameter $k$ can be expressed as additive combinations of complete elliptic integrals and therefore as additive combinations of $\E$ and $\J_1$:
\begin{equation}
\frac{d\E}{dk}=\frac{1}{kk'^2}[\E-k\J_1] \qquad \frac{d\J_1}{dk}=\frac{1}{k'^2}[\E-k\J_1].
\end{equation}
Note the extremely useful property of the derivative
\begin{equation} \label{eq:2}
\frac{d\E}{d\J_1}=\frac{1}{k}.
\end{equation}

We need to derive non-linear differential equation for the function $\E(\E_1)$. For this purpose, we find the derivative of $\E_1$ with respect to modular parameter $k$
\begin{equation}
\frac{d\E_1}{dk}=\frac{d(\J_1-\E+1)}{dk}=\frac{k-1}{kk'^2}(\E-k\J_1)
\end{equation}
and notice that the derivative of energy $\E$ with respect to modified spin $\E_1$ again has the property to depend only on the modular parameter $k$
\begin{equation} \label{eq:3}
\frac{d\E}{d\E_1}=\frac{1}{k-1}.
\end{equation}
Proceed with the second derivative
\begin{align}
\frac{d^2\E}{d\E_1^2}&=\frac{d}{d\E_1}\left(\frac{d\E}{d\E_1}\right)=\left(\frac{d\E_1}{dk}\right)^{-1}\frac{d}{dk}\left(\frac{1}{k-1}\right)\nonumber\\&=\frac{k(k+1)}{(k-1)^2\left[\E-k(\E_1+\E-1)\right]}.
\end{align}
Using \eqref{eq:3} we obtain
\begin{equation} \label{eq:4}
\left[(\E_1-1)\frac{d\E}{d\E_1}+\E+\E_1-1\right]\frac{d^2\E}{d\E_1^2}+2\left(\frac{d\E}{d\E_1}\right)^3+3\left(\frac{d\E}{d\E_1}\right)^2+\frac{d\E}{d\E_1}=0.
\end{equation}

Motivated by the type of series $\E(\E_1)$ one can make the following ansatz for $\E_1\rightarrow 0$ as a solution of \eqref{eq:4}:
\begin{equation}
\E(\E_1)=f_c+f\ln(\E_1)+\sum_{n=1}^{\infty}\sum_{m=0}^{n}f_{nm}\E_1^n\ln^m\E_1.
\end{equation}
This way we obtain recurrence relations between the coefficients
\begin{align}
&f=-\frac{1}{2},\qquad f_{11}=\frac{1}{4},\qquad f_{10}=\frac{1}{8}(-3-4f_c),\nonumber\allowdisplaybreaks[4]\\
&f_{22}=\frac{1}{16},\qquad f_{21}=\frac{1}{16}(1-4f_c),\qquad f_{20}=\frac{1}{64}(1-8f_c+16f_c^2),\nonumber\allowdisplaybreaks[4]\\
&f_{33}=\frac{1}{24},\qquad f_{32}=\frac{1}{32}(3-8f_c),\qquad f_{31}=\frac{1}{32}(3-12f_c+16f_c^2),\nonumber\allowdisplaybreaks[4]\\
&f_{30}=\frac{1}{96}(3-18f_c+36f_c^2-32f_c^3),\nonumber\allowdisplaybreaks[4]\\
&f_{44}=\frac{5}{128},\qquad f_{43}=\frac{1}{96}(13-30f_c),\qquad f_{42}=\frac{1}{256}(53-208f_c+240f_c^2),\nonumber\allowdisplaybreaks[4]\\
&f_{41}=\frac{1}{256}(39-212f_c+416f_c^2-320f_c^3),\allowdisplaybreaks[4]\\
&f_{40}=\frac{1}{6144}(279-1872f_c+5088f_c^2-6656f_c^3+3840f_c^4),\nonumber\allowdisplaybreaks[4]\\
&f_{55}=\frac{7}{160},\qquad f_{54}=\frac{1}{768}(157-336f_c),\qquad f_{53}=\frac{1}{384}(163-628f_c+672f_c^2),\nonumber\allowdisplaybreaks[4]\\
&f_{52}=\frac{1}{256}(121-652f_c+1256f_c^2-896f_c^3),\nonumber\allowdisplaybreaks[4]\\
&f_{51}=\frac{1}{192}(54-363f_c+978f_c^2-1256f_c^3+672f_c^4),\nonumber\allowdisplaybreaks[4]\\
&f_{50}=\frac{1}{61440}(4347-34560f_c+116160f_c^2-208640f_c^3+200960f_c^4-86016f_c^5), \nonumber
& \cdots \nonumber
\end{align}
All the coefficients are nicely determined through $f_c$ alone.

\subsection{Inverse spin function}\label{sec:InvSpinFunc}

In this subsection we start with the expansion using Picard-Fuchs equation. Next we will proceed with the method suggested in \cite{Floratos:2013cia}.

\paragraph{Using Picard-Fuchs equations}
The non-linear differential equation for the dispersion relation $\mathcal{E}=\mathcal{E}(\mathcal{J}_1)$ is:
\begin{equation}
\left(\mathcal{E}(\mathcal{J}_1)\frac{d\mathcal{E}}{d\mathcal{J}_1}-\mathcal{J}_1\right)\frac{d^2\mathcal{E}}{d\mathcal{J}_1^2}+\left(\frac{d\mathcal{E}}{d\mathcal{J}_1}\right)^3-\frac{d\mathcal{E}}{d\mathcal{J}_1}=0.
\end{equation}
We will search for solution by making the following ansatz:
\begin{equation}
\mathcal{E}(\mathcal{J}_1)=\mathcal{J}_1+f_c+\sum_{n=1}^\infty\sum_{m=0}^{n-1}f_{nm}\mathcal{J}_1^m \left(e^{-2\mathcal{J}_1-2}\right)^n.
\end{equation}
We obtain the following coefficients which all depend on one undetermined coefficient:
\begin{align}\label{eq:coeff}
&f_c=1,\quad f_{21}=f_{10}^2,\quad f_{20}=-\frac{f_{10}^2}{4},\quad  f_{32}=2f_{10}^3,\quad f_{31}=-\frac{f_{10}^3}{2},\quad f_{30}=\frac{f_{10}^3}{2},\nonumber\\
&f_{43}=\frac{16f_{10}^4}{3},\quad f_{42}=-f_{10}^4,\quad f_{41}=\frac{19f_{10}^4}{8},\quad f_{40}=-\frac{21f_{10}^4}{64},\\
&f_{54}=\frac{50f_{10}^5}{3},\quad f_{53}=-\frac{5f_{10}^5}{3},\quad f_{52}=\frac{81f_{10}^5}{8},\quad f_{51}=-\frac{55f_{10}^5}{32},\quad f_{50}=\frac{93f_{10}^5}{128}.\nonumber
\end{align}
This way, we need to obtain the coefficient $f_{10}$ by dint of another independent method. Such method is introduced underneath and we will see that $f_{10}=-4$. In addition, all the coefficients \eqref{eq:coeff} are in perfect agreement with these ones determined by the method of inverse spin function.

\paragraph{Using Lambert function}In the rest of this subsection we will follow the method suggested in \cite{Floratos:2013cia} in order to invert the series of angular momentum $\mathcal{J}_1(x)$, namely $x=x(\mathcal{J}_1)$, and then by substituting $x=x(\mathcal{J}_1)$ into $\mathcal{E}(x)$ to obtain the series of the dispersion relation $\mathcal{E}(\mathcal{J}_1)$ up to some order. For the needs of the upcoming considerations let us write down the series \eqref{eq:energy1} and \eqref{eq:spin1} in more convenient form. The coefficients we use in this section do not have to be confused with the coefficients used in section \ref{sec:ExpLongFoldedStr}.
\begin{equation}
\mathcal{E}=\sqrt{1-x}\cdot\sum_{n=0}^\infty x^n(d_n\ln x+h_n)=-\sum_{n=0}^\infty x^n\cdot\sum_{k=0}^n\frac{(2k-3)!!}{(2k)!!}(d_{n-k}\ln x+h_{n-k})\label{eq:newEseries}
\end{equation}
\begin{equation}
\mathcal{J}_1=\sum_{n=0}^\infty x^n(c_n\ln x+b_n)\label{eq:newJseries}
\end{equation}
Here the series for the energy and angular momentum are written in a way that they look similar to each other. Each coefficient in formulas \eqref{eq:newEseries} and \eqref{eq:newJseries} has the following simple form:
\begin{equation}
d_n=-\frac{1}{2}\left(\frac{(2n-1)!!}{(2n)!!}\right)^2,\qquad h_n=-4d_n\cdot(\ln 2+H_n-H_{2n}),
\end{equation}
\begin{equation}
c_n=-\frac{d_n}{2n-1},\qquad b_n=-4c_n\cdot\left[\ln 2+H_n-H_{2n}+\frac{1}{2(2n-1)}\right],
\end{equation}
where $n=0,1,2,\ldots$ and $H_n$ are the harmonic numbers. We start with solving equation \eqref{eq:newJseries} for $\ln x$:
\begin{equation}\label{eq:lnseries}
\ln x=\left[\frac{\mathcal{J}_1-b_0}{c_0}-\sum_{n=1}^\infty\frac{b_n}{c_0}x^n\right]\cdot\sum_{n=0}^\infty(-1)^n\left(\sum_{k=1}^\infty c_kx^k\right)^n.
\end{equation}
The above equation looks very complicated but one can get rid of the logarithm by taking exponent and then \eqref{eq:lnseries} acquires the following simpler form:
\begin{equation}\label{eq:x0series}
x=x_0\cdot\exp{\sum_{n=1}^\infty a_nx^n}=x_0\cdot\exp{\left(a_1x+a_2x^2+a_3x^3+\ldots\right)},
\end{equation}
where
\begin{equation}
x_0\equiv\exp\left[\frac{\mathcal{J}_1-b_0}{c_0}\right]=16\,e^{-2\mathcal{J}_1-2}.
\end{equation}
The advantage of this transformation is that we have especially convenient series for $x_0$ and the coefficients $a_n$ can be easily determined to the necessary order from \eqref{eq:lnseries}. Note that $x_0$ does not depend on $x$ but only on the angular momentum $\mathcal{J}_1$. This fact allows to revert the series \eqref{eq:x0series} by making use of the Lagrange inversion theorem, thus obtaining series for the variable $x$ in terms of $x_0$. In our special case one can apply the Lagrange-B\"urmann formula in order to get the inverse function:
\begin{equation}
x=\sum_{n=1}^\infty\frac{x_0^n}{n!}\cdot\left\{\frac{d^{n-1}}{dz^{n-1}}\exp\left[\sum_{m=1}^\infty n\,a_mz^m\right]\right\}_{z=0}.
\end{equation}
Differentiating $n$ times and taking the limit $z\rightarrow 0$, we find out an explicit form of the inverse function
\begin{equation}\label{eq:xreversed}
x=\sum_{n=1}^\infty x_0^n\cdot\sum_{k,j_i=0}^{n-1}\frac{n^k}{n!}\binom{n-1}{j_1,j_2,\ldots ,j_{n-1}}a_1^{j_1}a_2^{j_2}\ldots a_{n-1}^{j_{n-1}},
\end{equation}
with the following two constraints on the powers of coefficients $a_i$ satisfied
\begin{equation}
j_1+j_2+\ldots+j_{n-1}=k\quad\&\quad j_1+2j_2+\ldots+(n-1)j_{n-1}=n-1.
\end{equation}
From equation \eqref{eq:lnseries} can be deduced that all the $a_i$'s are linear in $\mathcal{J}_1$ therefore the inverse spin function $x=x(\mathcal{J}_1)$ is obliged to accept the general form
\begin{equation}\label{eq:xJ1series}
x=\sum_{n=1}^\infty x_0^n\cdot\sum_{k=0}^{n-1}a_{nk}\mathcal{J}_1^k,
\end{equation}
where the constants $a_{nk}$ have to be determined from \eqref{eq:xreversed}. The fact that the highest degree of $\mathcal{J}_1$ is $k=n-1$ becomes transparent if we combine the two constraints on the values of $j_i$'s as follows:
\begin{equation}\label{eq:constraints}
\left.\begin{aligned}
j_1+j_2+\ldots+j_{n-1}&=k\\
j_1+2j_2+\ldots+(n-1)j_{n-1}&=n-1
\end{aligned}\right\}
\Rightarrow k+j_2+\ldots+(n-2)j_{n-1}=n-1.
\end{equation}
Another very important information can be extracted from the constraints \eqref{eq:constraints}. This information is essential for taking decision which of the coefficients $a_i$ make contributions to certain power of $\mathcal{J}_1$, i.e. which $a_i$ constitute the coefficient $a_{nk}$ for certain $k$. The rule is as follows. The coefficients $a_{n\,n-1}$ (leading terms) are formed by the leading in $\mathcal{J}_1$ terms of $a_1$, the coefficients $a_{n\,n-2}$ (subleading terms) are formed by $a_1$ and the leading in $\mathcal{J}_1$ terms of $a_2$ etc., thereby the coefficients $a_{n\,n-m}$ are formed by $a_1,\ldots ,a_{m-1}$ and the leading in $\mathcal{J}_1$ terms of $a_m$. This rule becomes obvious if one take some $j_m\neq 0$ (at least 1), then from $k+j_2+\ldots+(n-2)j_{n-1}=n-1$ follows that $k=j_m+\ldots +j_{n-1}\leq n-m$.

\subsection{Anomalous dimensions}
Let us now use the obtained general formula \eqref{eq:xreversed} for $x(\mathcal{J}_1)$ to calculate the anomalous scaling dimensions $\gamma=\mathcal{E}-\mathcal{J}_1$ of spinning in $\mathbb{CP}^3$, closed, folded strings as a function of $\mathcal{J}_1$. First we will write down the anomalous dimensions in the already well-known form
\begin{equation}\label{eq:E-Jseries}
\mathcal{E}-\mathcal{J}_1=\sum_{n=0}^\infty x^n(f_n\ln x+g_n)=\sum_{n=0}^\infty x^n\left[A_n+f_n\ln\frac{x}{x_0}\right],
\end{equation}
where the new coefficients are defined by
\begin{equation}
f_n\equiv -c_n-\sum_{k=0}^n\frac{(2k-3)!!}{(2k)!!}\cdot d_{n-k},\quad g_n\equiv -b_n-\sum_{k=0}^n\frac{(2k-3)!!}{(2k)!!}\cdot h_{n-k},\quad n=0,1,2,\ldots\,,
\end{equation}
and for convenience we introduce $x_0$ by means of the coefficients $A_n$:
\begin{equation}
A_n\equiv g_n+f_n\ln x_0=g_n+2f_n(2\ln 2-\mathcal{J}_1-1).
\end{equation}
The expansions of anomalous scaling dimensions $\gamma=\gamma(\mathcal{J}_1)$ and inverse spin function $x=x(\mathcal{J}_1)$ both contain same terms but with different coefficients in front of them:
\begin{align}
\text{Leading terms (L): }&\mathcal{J}_1^{n-1}\left(e^{-2\mathcal{J}_1-2}\right)^n\nonumber\\
\text{Next-to-Leading/Subleading terms (NL): }&\mathcal{J}_1^{n-2}\left(e^{-2\mathcal{J}_1-2}\right)^n\\
\text{Next-to-Next-to-Leading terms (NNL): }&\mathcal{J}_1^{n-3}\left(e^{-2\mathcal{J}_1-2}\right)^n\nonumber\\
&\vdots\nonumber
\end{align}
The series \eqref{eq:newEseries}, \eqref{eq:newJseries}, and \eqref{eq:E-Jseries} look very similar to each other and therefore if we want to derive $\mathcal{E}-\mathcal{J}_1$ up to a certain order we need to calculate the inverse spin function $x=x(\mathcal{J}_1)$ up to the same order. This conclusion can be easily seen if we rewrite \eqref{eq:x0series} and \eqref{eq:xJ1series} as follows:
\begin{equation}
\ln\frac{x}{x_0}=\sum_{k=1}^\infty a_kx^k=a_1x+a_2x^2+a_3x^3+\ldots ,
\end{equation}
\begin{equation}
x=\sum_{n=1}^\infty x_0^n\cdot\sum_{k=0}^{n-1}a_{nk}\mathcal{J}_1^k=\sum_{n=1}^\infty\mathcal{J}_1^{n-1} x_0^n\cdot\sum_{k=0}^{n-1}\frac{\tilde{a}_{nk}}{\mathcal{J}_1^k}=\frac{1}{\mathcal{J}_1}\sum_{n=1}^\infty\mathcal{J}_1^{n}x_0^n\cdot\sum_{k=0}^{n-1}\frac{\tilde{a}_{nk}}{\mathcal{J}_1^k},
\end{equation}
where we redefine the constants $a_{nk}=\tilde{a}_{n\,n-k-1}$ and as before $a_n$ are linear functions of $\mathcal{J}_1$. Finally, the anomalous dimensions \eqref{eq:E-Jseries} acquire the following, convenient for calculations, form:
\begin{equation}
\mathcal{E}-\mathcal{J}_1=\sum_{n=0}^\infty x^n\left[A_n+f_n\ln\frac{x}{x_0}\right]=\sum_{n=0}^\infty x^n\left[A_n+\sum_{k=1}^\infty f_n\,a_kx^k\right].
\end{equation}

\paragraph{Leading terms}
To see how the above considerations work, let us apply them in practice to work out the leading order terms in the large $\mathcal{J}_1$ expansion of anomalous dimensions which means to find the coefficients of series
\begin{equation}
E-J_1\bigg|_{(L)}=\sum_{n=1}^\infty\mathfrak{a}_n\mathcal{J}_1^{n-1}\left(e^{-2\mathcal{J}_1-2}\right)^n.
\end{equation}
For this purpose, we need to calculate the leading terms of $x$, namely all the constants $\alpha_n$ have to be computed:
\begin{equation}
x_{(L)}=\sum_{n=1}^\infty\alpha_n\mathcal{J}_1^{n-1}\left(e^{-2\mathcal{J}_1-2}\right)^n.
\end{equation}
To do so, we have to collect the coefficients that multiply $x^0=1$ on the right-hand side of equation \eqref{eq:lnseries} and only leading in $\mathcal{J}_1$ coefficients that multiply $x^1=x$. Thus equation \eqref{eq:lnseries} take the form:
\begin{equation}
\ln x_{(L)}=\frac{\mathcal{J}_1-b_0}{c_0}-\frac{c_1}{c_0^2}\mathcal{J}_1\cdot x_{(L)}\Rightarrow x_0=x_{(L)}\exp\left[\frac{c_1}{c_0^2}\mathcal{J}_1\cdot x_{(L)}\right]=x_{(L)}\,e^{\mathcal{J}_1\cdot x_{(L)}/2},
\end{equation}
where $x_0=16\,e^{-2\mathcal{J}_1-2}$. We can reverse the above function either by making use of the Lagrange inversion theorem or just by employing formula \eqref{eq:iterexp} for the following iterated exponentiation:
\begin{equation}
x_{(L)}=x_0\,e^{-x_0\mathcal{J}_1/2\cdot e^{-x_0\mathcal{J}_1/2\cdot e^{\iddots}}}=\frac{2}{\mathcal{J}_1}W\left(8\mathcal{J}_1\,e^{-2\mathcal{J}_1-2}\right)=\sum_{n=1}^\infty \alpha_n\mathcal{J}_1^{n-1}\left(e^{-2\mathcal{J}_1-2}\right)^n,
\end{equation}
where we have chosen the principal branch\footnote{The principal branch provides the correct limiting behavior of $x$, i.e. $x\rightarrow 0^+$ when $\mathcal{J}_1\rightarrow+\infty$.} of Lambert $W$ function and have derived
\begin{equation}\label{eq:alphas}
\alpha_n\equiv(-1)^{n+1}2^{3n+1}\cdot\frac{n^{n-1}}{n!}.
\end{equation}
The final step in obtaining the large angular momentum expansion of anomalous dimensions $E-J_1$ is to put $x_{(L)}$ into \eqref{eq:E-Jseries} and then to retain only leading in $\mathcal{J}_1$ terms. Thus we end up with the following series:
\begin{align}
E-J_1\bigg|_{(L)}&=\frac{\sqrt{\tilde{\lambda}}}{2\pi}\left[1+g_1x_{(L)}-2f_2\mathcal{J}_1x_{(L)}^2\right]= \frac{\sqrt{\tilde{\lambda}}}{2\pi}\left[1-\frac{x_{(L)}}{4}-\frac{\mathcal{J}_1x_{(L)}^2}{16}\right]\nonumber\\
&=\frac{\sqrt{\tilde{\lambda}}}{2\pi}\left\{1-\frac{1}{4\mathcal{J}_1}\left[2\,W\left(8\mathcal{J}_1\, e^{-2\mathcal{J}_1-2}\right)+W^2\left(8\mathcal{J}_1\,e^{-2\mathcal{J}_1-2}\right)\right]\right\}\nonumber\\
&=\frac{\sqrt{\tilde{\lambda}}}{2\pi}\left\{1-\frac{1}{16}\sum_{n=1}^\infty\left[4\alpha_n+\sum_{k=1}^{n-1}\alpha_k\alpha_{n-k}\right]\cdot\mathcal{J}_1^{n-1}\left(e^{-2\mathcal{J}_1-2}\right)^n\right\}.
\end{align}

\paragraph{Next-to-leading terms}
We will sketch in brief the derivation of next-to-leading terms in the large $\mathcal{J}_1$ expansion of anomalous dimensions, i.e. we will calculate the coefficients in the series
\begin{equation}
E-J_1\bigg|_{(NL)}=\sum_{n=2}^\infty\mathfrak{b}_n\mathcal{J}_1^{n-2}\left(e^{-2\mathcal{J}_1-2}\right)^n.
\end{equation}
To accomplish that, in addition to the leading terms one should compute also the next-to-leading terms of $x$ in equation \eqref{eq:xreversed}, more precisely:
\begin{equation}
x_{(NL)}=\sum_{n=2}^\infty\beta_n\mathcal{J}_1^{n-2}\left(e^{-2\mathcal{J}_1-2}\right)^n.
\end{equation}
Following again the already familiar strategy, one has to collect the coefficients in front of $x^0$, $x^1$ on the right-hand side of \eqref{eq:lnseries} and just the leading in $\mathcal{J}_1$ coefficients in front of $x^2$. In consequence equation \eqref{eq:lnseries}, with accuracy up to subleading order, acquires the form:
\begin{gather}
\ln x_{(L+NL+\ldots)}=\frac{\mathcal{J}_1-b_0}{c_0}-\frac{\mathcal{J}_1c_1+b_1c_0-b_0c_1}{c_0^2}x_{(L+NL+\ldots)}+\frac{c_1^2-c_0c_2}{c_0^3}\mathcal{J}_1x_{(L+NL+\ldots)}^2\nonumber\\
\Rightarrow x_{(L+NL+\ldots)}=x_0\exp\left[-\frac{\mathcal{J}_1+1}{2}x_{(L+NL+\ldots)}-\frac{7\mathcal{J}_1}{32}x_{(L+NL+\ldots)}^2\right].
\end{gather}
The next step is to invert the above equation for $x_{(L+NL+\ldots)}$ by making use of the Lagrange-B\"urmann formula:
\begin{equation}
x_{(L+NL+\ldots)}=\sum_{n=1}^\infty\frac{x_0^n}{n!}\sum_{\substack{k,j_1=0\\n-1=k+j_1\\0\leq j_1\leq k}}^{n-1}(-1)^kn^k\frac{(n-1)!}{(k-j_1)!j_1!}\left(\frac{\mathcal{J}_1+1}{2}\right)^{k-j_1}\left(\frac{7\mathcal{J}_1}{32}\right)^{j_1}.
\end{equation}
In the above series we need to separate only the subleading $\mathcal{J}_1$-terms since we have already obtained the leading terms in the previous paragraph. After expansion of the binomial and careful selection of the terms, we find
\begin{equation}
x_{(NL)}=\sum_{n=1}^\infty\frac{x_0^n}{n!}\left\{(-1){n-1}n^{n-1}\frac{n-1}{2^{n-1}}+(-1)^{n-2}n^{n-2}\frac{7(n-1)(n-2)}{2^{n+2}}\right\}\mathcal{J}_1^{n-2},
\end{equation}
which allows us to write down $x$ up to subleading order in the compact form
\begin{equation}\label{eq:xLNL}
x_{(L+NL)}=\sum_{n=1}^\infty\left(\alpha_n\mathcal{J}_1^{n-1}+\beta_n\mathcal{J}_1^{n-2}\right)\cdot \left(e^{-2\mathcal{J}_1-2}\right)^n,
\end{equation}
where the $\alpha_n$'s have already been defined by \eqref{eq:alphas} and $\beta_n$'s are defined by:
\begin{equation}
\beta_n\equiv(-1)^{n+1}2^{3n-2}\frac{n^{n-2}}{n!}(n-1)(n+14).
\end{equation}
Now formulas \eqref{eq:Wfirst}-\eqref{eq:Wlast} can be employed in order to represent the series \eqref{eq:xLNL} in terms of Lambert $W$ function with argument $W(8\mathcal{J}_1e^{-2\mathcal{J}_1-2})$:
\begin{equation}\label{eq:xLNLW}
x_{(L+NL)}=\frac{2}{\mathcal{J}_1}W-\frac{1}{4\mathcal{J}_1^2}\frac{W^2(7W+8)}{1+W}.
\end{equation}
The last thing we have to do is to put \eqref{eq:xLNLW} into \eqref{eq:E-Jseries} an retain only the leading and subleading terms. Since we have already obtained the leading terms, here we isolate only the subleading terms which lead us to the final result:
\begin{align}
E-J_1\bigg|_{(NL)}=-\frac{\sqrt{\tilde{\lambda}}}{128\pi}\sum_{n=1}^\infty\left\{\vphantom{\sum_{k,m=1}^{n-2}}\right.&16\beta_n+\sum_{k=1}^{n-1}\alpha_k\left[9\alpha_{n-k}+8\beta_{n-k}\right]\nonumber\\
& \left.{}+4\sum_{k,m=1}^{n-2}\alpha_k\alpha_m\alpha_{n-k-m}\right\}\cdot\mathcal{J}_1^{n-2}\left(e^{-2\mathcal{J}_1-2}\right)^n.
\end{align}

As mentioned above, the results of the three methods are consistent. Expressing the anomalous dimensions in different form may be helpful studying different properties of the ABJM theory. We will go back to these issues in a forthcoming publication.

\section{Conclusions}

According to the AdS/CFT correspondence, the anomalous dimensions of the gauge theory operators are given by the
dispersion relation of their dual $AdS_4\times \mathbb{CP}^3$ strings.
We have computed the large-spin expansion of anomalous dimensions of gauge theory operators in ABJM theory using results from string theory side. For the simple folded string solutions of \cite{Chen:2008qq} the energy and momenta are expressed in terms of elliptic integrals and therefore, to obtain the desired dispersion relations one has to invert the elliptic integrals and solve for the energy in terms of momenta. The inversion of elliptic integrals with respect to the modular parameter $\kappa$, is not an easy task.

We consider two types of folded string solutions, short and long, characterized by the modular parameters of the elliptic functions.  Due the Legendre relation that connects complete elliptic integrals of the first and second kind \eqref{eq:Legendre}, there is a remarkable duality between short and long strings \eqref{eq:EJduality}. According to this formula for each solution of energy $E$ and spin $J_1$, there exists a dual solution whose energy $E'$ and spin $J'_1$ with modular parameters related by $k^2+k'^2=1$. In terms of the anomalous dimensions this duality has the form \eqref{eq:Egammaduality}. We presented in details the considerations for the simplest case of folded strings given in \cite{Chen:2008qq}. The results for the more complicated case given in that paper contain long and not too informative expressions. They  are collected in Appendix C.

We found expressions for the dispersion relations in the form of logarithmic power series and exponential power series in momentum $\mathcal{J}_1$ for leading and first subleading terms (using Lambert functions). We checked that the three approaches we used are consistent giving same results. It is interesting to note that logarithmic expansions are typical for the expansions in AdS part of the geometry (see for instance \cite{Floratos:2013cia}). In $\mathbb{CP}^3$ case the expressions for the energy and the momenta in terms of elliptic integrals is very similar to the $AdS_5$ case which argues why this happens in our case.

The results in this note can be extended to include finite size corrections. It would be interesting to pursue the idea to look for recurrent relations allowing to obtain subleading contributions. Clues for that may come from different approaches we used to obtain the dispersion relations, especially Picard-Fuchs equation. Promotion of these considerations to quantum level is also a challenging task.

\section*{Acknowledgments}
The authors would like to thank D.~Arnaudov and T.~Vetsov for fruitful discussions. This work was supported in part by the Austrian Science Fund (FWF) project I 1030-N16.

\begin{appendix}

\section{Elliptic integrals of first and second kind}\label{elliptic}

The defining relations are
\begin{align}
& \bK=\int\limits_0^1\frac{dx}{\sqrt{(1-x^2)(1-k^2x^2)}}\allowdisplaybreaks[4]\\
& \bE=\int\limits_0^1\frac{dx\sqrt{1-k^2x^2}}{\sqrt{(1-x^2)}}.
\end{align}
One can pass however, to the Abelian periods known from the algebraic geometry
\begin{equation}
I_1=\int\limits_0^1\frac{dz}{\sqrt{P(z)}}, \quad I_2=\int\limits_0^1\frac{z\,dz}{\sqrt{P(z)}},
\end{equation}
where $P(z)=z(1-z)(1-k^2z)$. Then under the identification $z=x^2$ we find the relations
\begin{equation}
\bK=\frac{1}{2}I_1, \qquad \bE=\frac{1}{2}I_1-\frac{k^2}{2}I_2.
\end{equation}
The conclusion is that the elliptic functions are additive combinations of the Abelian periods!

One can use the dependence on the modular parameter $k$ ($k^{\prime 2}=1-k^2$) and differentiate the
Abelian periods using their defining equations:
\begin{equation}
\frac{d I_1}{dk}=\frac{k}{k^\prime}(I_1-I_2), \qquad \frac{d I_2}{dk}=
\frac{1}{kk^{\prime 2}}(I_1-(2-k^2)I_2).
\end{equation}
and the corresponding equations for the complete elliptic integrals
\begin{equation}
\frac{d\bK}{dk}=\frac{1}{kk^{\prime 2}}(\bE-k^{\prime 2}\bK), \quad \frac{d\bE}{dk}=
\frac{1}{k}(\bE-\bK).
\end{equation}

There is a well known expansion of the elliptic integral of first kind \bK(k) for the small values of the module $k^2<1$
\begin{equation}\label{eq:expansionKsmall}
\bK(k)=\frac{\pi}{2}\left\{1+\sum_{n=1}^{\infty}\left[\frac{(2n-1)!!}{(2n)!!}\right]^2 k^{2n}\right\}, \qquad k^2<1
\end{equation}
and for the large values $1-k^2<1$
\begin{align} \label{eq:expansionK}
\bK(k)&=-\frac{1}{2\pi}\sum_{n=0}^{\infty}\left[\frac{\G(n+1/2)}{n!}\right]^2(1-k^2)^n\cdot\nonumber\\
&\cdot\left[\ln(1-k^2)+2\psi(n+1/2)-2\psi(n+1)\right],
\end{align}
where $\psi(z)$ is the digamma function. For the elliptic integral of second kind $\bE(k)$ the expansion for the small values of the module $k^2<1$ is
\begin{equation}\label{eq:expansionEsmall}
\bE(k)=\frac{\pi}{2}\left\{1-\sum_{n=1}^{\infty}\left[\frac{(2n-1)!!}{(2n)!!}\right]^2 \frac{k^{2n}}{2n-1}\right\}, \qquad k^2<1
\end{equation}
and for the large values $1-k^2<1$ it is
\begin{align} \label{eq:expansionE}
\bE(k)&=1-\frac{(1-k^2)}{2\pi}\sum_{n=0}^{\infty}\frac{\G(n+1/2)\G(n+3/2)}{n!(n+1)!}(1-k^2)^n\cdot\nonumber\\
&\cdot\left[\ln(1-k^2)+\psi(n+1/2)+\psi(n+3/2)-\psi(n+1)-\psi(n+2)\right].
\end{align}

\section{Lambert $W$ function}
\begin{figure}[h]
\centering
\includegraphics[width=0.6\textwidth]{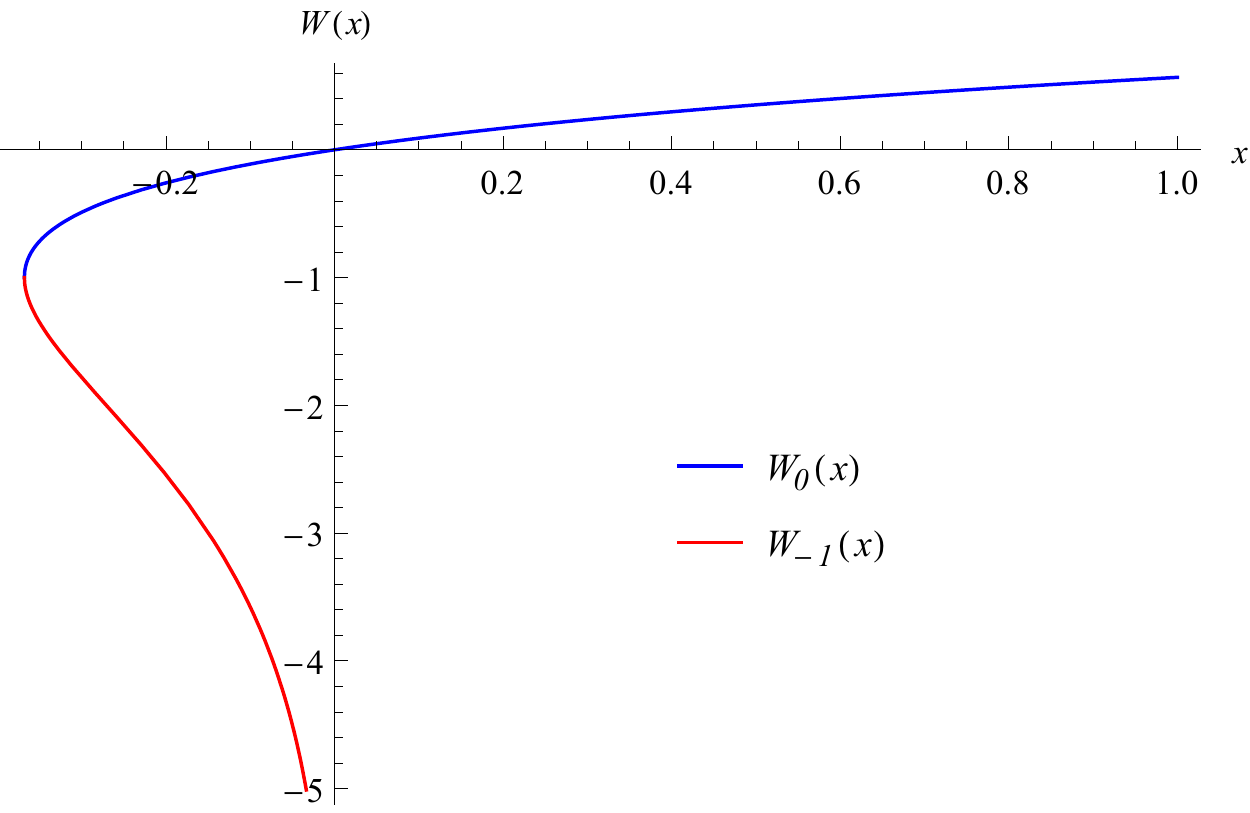}
\caption{The two real branches of the Lambert's $W$ function.}
\label{fig:LambertWPlot}
\end{figure}

The Lambert $W$ function is a multivalued function defined as a solution to the following transcendental equation:
\begin{equation}\label{eq:WDef}
W(z)e^{W(z)}=z
\end{equation}
for any complex number $z$. $W(x)$ possesses two real branches, expressly $W_0(x)$ in $[-e^{-1},\infty)$ and $W_{-1}(x)$ in $[-e^{-1},0]$, which are plotted on figure \ref{fig:LambertWPlot}. The point, at which the two branches are connected, is $(-e^{-1},-1)$. Both branches have well-defined Taylor series at $x=0$ \cite{Corless}:
\begin{equation}
W_0(x)=\sum_{n=0}^\infty(-1)^{n+1}\frac{(n+1)^n}{(n+1)!}x^{n+1}=\sum_{n=1}^\infty(-1)^{n+1}\frac{n^{n-1}}{n!}x^n,\quad|x|\leq e^{-1},
\end{equation}
\begin{equation}
W_{-1}(x)=\ln|x|-\ln\ln|x|+\sum_{n=0}^\infty\sum_{m=1}^\infty\frac{(-1)^n}{m!}\stirlingone{n+m}{n+1}(\ln|x|)^{-n-m}(\ln\ln|x|)^m,
\end{equation}
where the unsigned Stirling numbers of the first kind are denoted by $\stirlingone{n}{k}$. They count the number of permutations of $n$ elements which contain exactly $k$ cycles and can be computed recursively by:
\begin{equation}
\stirlingone{n}{k}=\stirlingone{n-1}{k-1}+(n-1)\stirlingone{n-1}{k},\quad \stirlingone{n}{0}=\stirlingone{0}{k}=0,\quad \stirlingone{0}{0}=1,\quad n,k\geq 1.
\end{equation}
The unsigned Stirling numbers of the first kind are related to the harmonic numbers $H_{n-1}$ and generalized harmonic numbers $H_{n-1}^{(2)}$ by:
\begin{equation}
\stirlingone{n}{1}=(n-1)!,\quad \stirlingone{n}{2}=(n-1)!H_{n-1},\quad \stirlingone{n}{3}=\frac{1}{2}(n-1)!\left[H_{n-1}^2-H_{n-1}^{(2)}\right].
\end{equation}

The W function also gives a solution to the problem of iterated exponentiation $x^{x^{x^{\iddots}}}$. Euler was the first to prove that this iteration converges for real $x$ between $e^{-e}$ and $e^{1/e}$ and the limit is
\begin{equation}\label{eq:iterexp}
x^{x^{x^{\iddots}}}=\frac{W(-\ln x)}{-\ln x}.
\end{equation}

Using the defining equation \eqref{eq:WDef} one can easily find out expressions for the derivatives and antiderivatives of Lambert's function. Here we list some of the most helpful formulas including derivatives and integrals of the $W_0$ function (the so-called principal branch):
\begin{equation}\label{eq:Wfirst}
W'(x)=\frac{W(x)}{x\left(1+W(x)\right)}
\end{equation}
\begin{equation}
xW'(x)=\sum_{n=1}^\infty(-1)^{n+1}\frac{n^n}{n!}x^n=\frac{W(x)}{1+W(x)}
\end{equation}
\begin{equation}
x\left(xW'(x)\right)'=\sum_{n=1}^\infty(-1)^{n+1}\frac{n^{n+1}}{n!}x^n=\frac{W(x)}{\left(1+W(x)\right)^3}
\end{equation}
\begin{equation}
\int W(x)\,dx=x\left(W(x)-1+\frac{1}{W(x)}\right)
\end{equation}
\begin{equation}
\int\frac{W(x)}{x}\,dx=\sum_{n=1}^\infty(-1)^{n+1}\frac{n^{n-2}}{n!}x^n=W(x)+\frac{W^2(x)}{2}
\end{equation}
\begin{equation}\label{eq:Wlast}
\int\frac{1}{x}\int\frac{W(x)}{x}\,dx^2=\sum_{n=1}^\infty(-1)^{n+1}\frac{n^{n-3}}{n!}x^n=W(x)+\frac{3W^2(x)}{4}+\frac{W^3(x)}{6}.
\end{equation}

\section{Folded strngs II}

There exists another folded string solution \cite{Chen:2008qq} that can be obtained if we fix $\xi=\pi/4$. In this case, the only possible way to have nontrivial solution is to require $\theta_1=\pm\theta_2$. From the equations of motion for $\theta_1$ and $\theta_2$ follow that this is only possible for $\omega_2=-\omega_3$. Explicitly, the equation of motion for $\theta_1$ has the form
\begin{equation}
\theta_1^{\prime\prime}=\omega_1\omega_2\sin\theta_1
\end{equation}
and it is consistent with the Virasoro constraint
\begin{equation}
\kappa^2=\theta_1^{\prime2}+\omega_1^2+\omega_2^2+2\omega_1\omega_2\cos\theta_1.
\end{equation}
In the same manner as for the simpler folded string solution, let $\theta_1(0)$ be the maximal value of $\theta_1$. Since $\theta_1$ is periodic function of $\sigma$, we have
\begin{equation}
-\theta_1(0)\leq\theta_1(\sigma)\leq\theta_1(0),\quad \kappa^2=\omega_1^2+\omega_2^2+2\omega_1\omega_2\cos\theta_1(0).
\end{equation}
There are two different cases here according to the sign of $\omega_1\omega_2$. If $\omega_1\omega_2>0$, $\theta_1$ is varying around $\pi$. On the contrary, if $\omega_1\omega_2<0$, the folded string is centered at $\theta_1=0$. Without loss of generality, we will consider the latter case, i.e. we will assume $\omega_1>0$ and $\omega_2<0$. Now it is easy to integrate the Virasoro constraint
\begin{equation}
2\pi=\int_0^{2\pi}d\sigma=\frac{2}{\sqrt{-\omega_1\omega_2}}\int_0^{\theta_1(0)}\frac{d\theta_1}{\sqrt{\sin^2\frac{\theta_1(0)}{2}-\sin^2\frac{\theta_1}{2}}}.
\end{equation}
This way we get
\begin{equation}
\sqrt{-\omega_1\omega_2}=\frac{2}{\pi}\mathbb{K}(k),
\end{equation}
where $k=\sin\frac{\theta_1(0)}{2}$. The energy of the folded string is just $E=\frac{1}{4}\sqrt{\tilde{\lambda}}\kappa$. The angular momenta are:
\begin{align}
&J_1=\frac{\sqrt{\tilde{\lambda}}}{2\pi\sqrt{-\omega_1\omega_2}}\left((\omega_1-\omega_2)\mathbb{K}(k)+2\omega_2\mathbb{E}(k)\right),\\
&J_2=\frac{\sqrt{\tilde{\lambda}}}{4\pi\sqrt{-\omega_1\omega_2}}\left((\omega_2-\omega_1)\mathbb{K}(k)+2\omega_1\mathbb{E}(k)\right),\\
&J_3=-J_2.
\end{align}
One can show that the following relation is satisfied
\begin{equation}
\frac{\omega_1}{2}J_1-\omega_2J_2=\frac{\sqrt{\tilde{\lambda}}}{8}\left(\omega_1^2-\omega_2^2\right).
\end{equation}
In terms of the convenient quantities $\mathcal{E}=\frac{E}{\sqrt{\tilde{\lambda}}},~\mathcal{J}_i=\frac{J_i}{\sqrt{\tilde{\lambda}}},~i=1,2,3$, we have
\begin{align}
&\omega_1=\mathbb{K}(k)\frac{\mathbb{K}(k)\mathcal{J}_1-2\mathcal{J}_2\left(2\mathbb{E}(k)-\mathbb{K}(k)\right)}{\mathbb{E}(k)\left(\mathbb{K}(k)-\mathbb{E}(k)\right)},\\
&\omega_2=\mathbb{K}(k)\frac{\mathbb{K}(k)\mathcal{J}_2-\mathcal{J}_1\left(2\mathbb{E}(k)-\mathbb{K}(k)\right)}{\mathbb{E}(k)\left(\mathbb{K}(k)-\mathbb{E}(k)\right)},
\end{align}
and the following key relations
\begin{align}
\left(\frac{\mathcal{E}}{\mathbb{K}(k)}\right)^2-\left(\frac{\mathcal{J}_1+2\mathcal{J}_2}{\mathbb{E}(k)}\right)^2&=\frac{4}{\pi^2}k^2,\label{eq:relE}\\
\left(\frac{\mathcal{J}_1-2\mathcal{J}_2}{\mathbb{E}(k)-\mathbb{K}(k)}\right)^2-\left(\frac{\mathcal{J}_1+2\mathcal{J}_2}{\mathbb{E}(k)}\right)^2&=\frac{4}{\pi^2}\label{eq:relJ}.
\end{align}
The above relations constitute complicated system in a manner that it looks hard to exclude modulus parameter $k$ in order to obtain some kind of dispersion relation. However, using brute force one can derive such dispersion relation at least for the small values of $k$. To do this, let us introduce new variables, namely $\mathcal{J}_+=\mathcal{J}_1+2\mathcal{J}_2$ and $\mathcal{J}_-=\mathcal{J}_1-2\mathcal{J}_2$. In terms of the new variables \eqref{eq:relE} and \eqref{eq:relJ} get the form
\begin{align}
\mathcal{E}^2&=\left(\frac{4}{\pi^2}k^2+\frac{\mathcal{J}_+^2}{\mathbb{E}^2(k)}\right)\mathbb{K}^2(k)\label{eq:Enew}\\
\mathcal{J}_-^2&=\left(\frac{4}{\pi^2}+\frac{\mathcal{J}_+^2}{\mathbb{E}^2(k)}\right)\left(\mathbb{E}(k)-\mathbb{K}(k)\right)^2\label{eq:Jminusnew}.
\end{align}
The main advantage here is that the expansions of the complete elliptic integrals are represented by power series for small values of $k$. It is laborious task to deal with power series, but still quite straightforward. Hence we will cling to the following procedure. First, we expand the complete elliptic integrals using \eqref{eq:expansionKsmall} and \eqref{eq:expansionEsmall}. Second, we obtain power series for $\mathcal{E}$ and $\mathcal{J}_-$ in terms of the variable $x=k^2$. Third, we fix $\mathcal{J}_+$ in \eqref{eq:Jminusnew} and reverse the series for $x$. The result for the inverse spin function $x=x(\mathcal{J}_+,\mathcal{J}_-)$ is:
\begin{equation}\label{eq:xSeries}
x=\frac{2\mathcal{J}_-}{\sqrt{1+\mathcal{J}_+^2}}+\frac{\left(-3-5\mathcal{J}_+^2\right)\mathcal{J}_-^2}{2(1+\mathcal{J}_+^2)^2}+\frac{\left(3+15\mathcal{J}_+^2+22\mathcal{J}_+^4\right)\mathcal{J}_-^3}{8(1+\mathcal{J}_+^2)^{7/2}}+\mathcal{O}(\mathcal{J}_-^4)
\end{equation}
Now we have to insert \eqref{eq:xSeries} in the expansion of \eqref{eq:Enew} this way getting dispersion relation in powers of $\mathcal{J}_-$:
\begin{equation}
\mathcal{E}=\mathcal{J}_++\frac{\sqrt{1+\mathcal{J}_+^2}\,\mathcal{J}_-}{\mathcal{J}_+}-\frac{\left(2+3\mathcal{J}_+^2\right)\mathcal{J}_-^2}{4\mathcal{J}_+^3\left(1+\mathcal{J}_+^2\right)}+\frac{\left(8+28\mathcal{J}_+^2+33\mathcal{J}_+^4+ 15\mathcal{J}_+^6\right)\mathcal{J}_-^3}{16\mathcal{J}_+^5\left(1+\mathcal{J}_+^2\right)^{5/2}}+\mathcal{O}(\mathcal{J}_-^4).
\end{equation}
Finally, one can get back to the original angular momenta $\mathcal{J}_1$ and $\mathcal{J}_2$:
\begin{align}
\mathcal{E}&=\mathcal{J}_1+2\mathcal{J}_2+\frac{\sqrt{1+(\mathcal{J}_1+2\mathcal{J}_2)^2}\,(\mathcal{J}_1-2\mathcal{J}_2)}{\mathcal{J}_1+2\mathcal{J}_2}-\frac{\left(2+3(\mathcal{J}_1+2\mathcal{J}_2)^2\right)(\mathcal{J}_1-2\mathcal{J}_2)^2}{4(\mathcal{J}_1+2\mathcal{J}_2)^3\left(1+(\mathcal{J}_1+2\mathcal{J}_2)^2\right)}\nonumber\\
&+\frac{\left(8+28(\mathcal{J}_1+2\mathcal{J}_2)^2+33(\mathcal{J}_1+2\mathcal{J}_2)^4+ 15(\mathcal{J}_1+2\mathcal{J}_2)^6\right)(\mathcal{J}_1-2\mathcal{J}_2)^3}{16(\mathcal{J}_1+2\mathcal{J}_2)^5\left(1+(\mathcal{J}_1+2\mathcal{J}_2)^2\right)^{5/2}}+\mathcal{O}(\mathcal{J}_-^4).
\end{align}

AdS/CFT dictionary provides correspondence between the conserved charges on string theory side and the composite operators on gauge theory side.
The identification of the operators corresponding to the above angular momenta was suggested in \cite{Chen:2008qq}
\eq{
\left\{
\begin{array}{ll}
{\rm Tr}\left((A_1B_1)^{\frac{J_{\psi}}{2}+J_{\varphi_1}}(A_2B_2)^{\frac{J_{\psi}}{2}-
J_{\varphi_1}}\right),&\hspace{3ex}\mbox{for $\omega_1+\omega_2>0$}\,,\\
{\rm Tr}\left((B_1^\dagger A_1^\dagger)^{-(\frac{J_{\psi}}{2}+J_{\varphi_1})}(A_2B_2)^{\frac{J_{\psi}}{2}-
J_{\varphi_1}}\right),&\hspace{3ex}\mbox{for $\omega_1+\omega_2<0$}\,.
\end{array}
\right.
}
Note that, if one takes $\omega_1=-\omega_2$, the above operators reduce to 
${\rm Tr}(A_2B_2)^{J_{\psi}}$, which is a BPS primary and has $\Delta=J_{\psi}$ without
 quantum correction, which is in disagreement with the string calculation.
 Since the dispersion relations \eqref{eq:relE} and \eqref{eq:relJ} are the same as
in the $AdS_5\times S^5$ case, it is natural to accept that point of view, namely, that there exists an interpolating function \cite{Gromov:2008qe} governing the anomalous dimension transition from $g^2$ in the weak coupling regime \cite{Minahan:2008hf}-\cite{Leoni:2010tb} to $g$ in the strong coupling one \cite{Nishioka:2008gz,Gaiotto:2008cg}.
\end{appendix}

\end{document}